\documentclass{appolb}
\usepackage{epsfig}
\usepackage{color}
\def\be{\begin{equation}}
\def\ee{\end{equation}}
\def\bea{\begin{eqnarray}}
\def\eea{\end{eqnarray}}
\definecolor{blueish}{rgb}{0.0,0.0,0.7}
\definecolor{greenish}{rgb}{0.0,0.7,0.0}
\definecolor{darkgreen}{rgb}{0.0,0.4,0.0}
\definecolor{turqoise}{rgb}{0.0,0.5,0.5}
\definecolor{gold}{rgb}{0.7,0.6,0.0}

\newcommand{\rse}{\mathcal{R}}
\newcommand{\bes}[1]{j^{#1}_{L_{#1}}}
\newcommand{\han}[1]{h^{(1)#1}_{L_{#1}}}
\newcommand{\pot}[2]{\mathcal{V}_{#1#2}^{(L_{#1},L_{#2})}}
\newcommand{\tmat}[2]{\mathcal{T}_{#1#2}^{(L_{#1},L_{#2})}}
\newcommand{\One}{1\!\!1}

\begin{document}
\title{Complex masses in the $\mathcal{S}$-matrix
\thanks{Talk at Workshop \em ``Excited QCD 2010'', \em Stara Lesn\'{a}, Slovakia,
31 Jan.\ -- 6 Feb.\ 2010}
}
\author{George Rupp$^*$, Susana Coito
\address{
Centro de F\'{\i}sica das Interac\c{c}\~{o}es Fundamentais,
Instituto Superior T\'{e}cnico, Technical University of
Lisbon, P-1049-001 Lisboa, Portugal} 
\and
Eef van Beveren
\address{
Centro de F\'{\i}sica Computacional, Departamento de F\'{\i}sica,
Universidade de Coimbra, P-3004-516 Coimbra, Portugal}
}
\maketitle
\begin{abstract}
Most excited hadrons have multiparticle strong decay modes, which can often
be described as resulting from intermediate states containing one or two
resonances. In a theoretical approach, such a description in terms of
quasi-two-particle initial and final states leads to unitarity violations,
because of the complex masses of the involved resonances. In the present 
paper, an empirical algebraic procedure is presented to restore unitarity
of the $\mathcal{S}$-matrix while preserving its symmetry. Preliminary
results are presented in a first application to $S$-wave $\pi\pi$ scattering,
in the framework of the Resonance-Spectrum Expansion.
\end{abstract}

\section{Introduction} 
Inspection of the Particle Data Group (PDG) tables \cite{PLB667p1} reveals that
most excited mesons and baryons have strong decay modes involving more than
two lighter hadrons. Moreover, one also verifies that many of these decays
can be considered ``cascades'' of decays from intermediate states involving
one or more resonances. A few mesonic examples of such processes are
\cite{PLB667p1}
\begin{enumerate}
\item
$f_0(1370)\;\to\;\rho\rho\,,\,2(\pi\pi)_{\mbox{\scriptsize $S$-wave}}\,,
\,\ldots\;\to\;4\pi\;;$ \\[-5mm]
\item
$K_2(1770)\;\to\;K_2^\ast(1430)\,\pi\,,\,Kf_2(1270)\,,\,\ldots\;\to\;
K\pi\pi\;;$\\[-5mm]
\item
$\phi(2170)\;\to\;\phi f_0(980)\;\to\;\phi\pi\pi\,,\,KK f_0(980)\;\to\;
KK\pi\pi\;.$
\end{enumerate}
In the first example, $2(\pi\pi)_{\mbox{\scriptsize $S$-wave}}$ stands for
$\sigma\sigma$, where $\sigma$ is the very broad $f_0(600)$ \cite{PLB667p1}
scalar resonance. In all three cases, the final state contains 3 or even
4 mesons that are stable with respect to strong interactions. In order to
describe such processes in a mathematically rigorous way, respecting unitarity,
one would have to resort to (relativistic) Faddeev \cite{SPJETP12p1014} or
Yakubovsky \cite{SJNP5p937} equations, respectively, in the scattering regime.
As a matter of fact, such a Faddeev approach was applied \cite{PRD78p074031}
to the $\phi(2170)$, as a three-body resonance in $\phi KK$ and
$\phi\pi\pi$, though not including the $KKf_0(980)$ component and also not
the 4-body $KK\pi\pi$ channel. Note that solving relativistic 3- or 4-body
equations with many channels is already a huge task, but it becomes absolutely
impracticable if one wants to go beyond an effective description as in
Ref.~\cite{PRD78p074031}, by taking into account the quark substructure of
the decaying resonance.

In experimental data analyses, one often falls back upon dispersive
or purely phenomenological parametrisations, as e.g.\ in
Refs.~\cite{EPJC52p55,NPB471p59} for the description of the $4\pi$ channel
in $S$-wave $\pi\pi$ scattering above 1~GeV and the $f_0(1370)$ resonance.

An alternative, theoretical approach is to interpret the intermediate state
in a cascade decay as a (quasi-)final state, containing one or two resonances
as outgoing particles. The problem then arises how to deal with the, in
principle, complex mass(es) of the resonance(s),
while preserving two-body unitarity. One way is by discretising the
resonance real-mass distribution(s), and accordingly including a large number
of channels having the corresponding threshold energies, with relative coupling
strengths given by the mass-distribution function. Something in this spirit 
has been carried out in a coupled-channel description of $S$-wave meson-meson
scattering in a chiral unitary model \cite{IJMPA24p581}. A drawback of such
a method is the proliferation of channels, besides the problem of properly
dealing with resonances not far above threshold.

In the present study, we shall try out a novel, intuitive approach, by simply
substituting complex values for the masses of the final-state resonances,
according to their listed \cite{PLB667p1} real (Breit-Wigner) masses and
widths. Although this step inexorably destroys unitarity, we can afterwards
restore it by a suitable redefinition of the $\mathcal{S}$-matrix. This quite
general procedure is then applied to a concrete test case, in the framework of
the Resonance-Spectrum Expansion. \\[-6mm]

\section{Resonance-Spectrum Expansion}
\begin{figure}[h]
\begin{tabular}{cc}
\epsfysize=45pt
\epsfbox{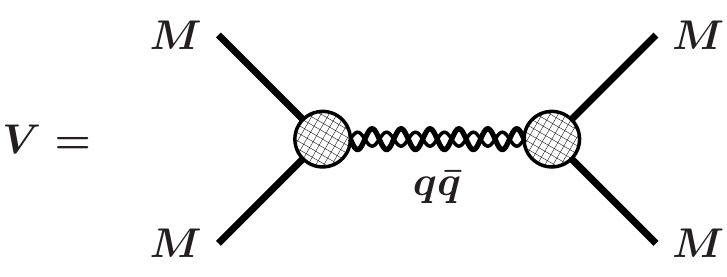}
&
\hspace*{5mm}
\epsfysize=45pt
\epsfbox{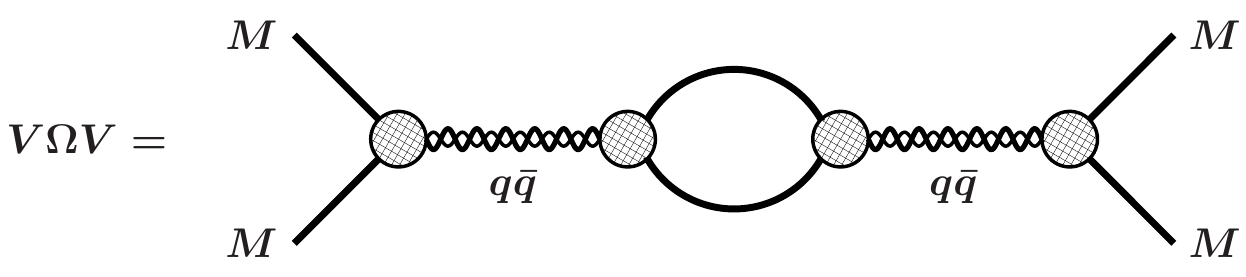} 
\end{tabular}
\caption{Born and one-loop term of the RSE effective meson-meson interaction.}
\label{bornloop}
\end{figure}

The Resonance-Spectrum Expansion (RSE) is a model for the scattering
of two mesons in non-exotic channels, via an infinite set of intermediate
$s$-channel
$q\bar{q}$ states, i.e., a kind of Regge propagators \cite{AOP324p1620}. The
confinement spectrum for these bare $q\bar{q}$ states can, in principle, be
chosen freely, but in all successful phenomenological applications so far we
have used a harmonic-oscillator (HO) spectrum with flavour-indepedent
frequency (see Ref.~\cite{AOP324p1620} for several references). Because of the
separability of the effective meson-meson interaction, the RSE model can be
solved in closed form. The relevant Born and one-loop diagrams are depicted in
Fig.~\ref{bornloop}.
For $N$ meson-meson channels and several $q\bar{q}$ channels, the effective
potential has the form
\begin{eqnarray}
\pot{i}{j}(p_i,p'_j;E) &=& \displaystyle\lambda^2r_0\,\bes{i}(p_ir_0)\,
\bes{j}(p'_jr_0)\,\sum_{\alpha=1}^{N_{q\bar{q}}}\sum_{n=0}^{\infty}
\displaystyle\frac{g_i^{(\alpha)}(n) g_j^{(\alpha)}(n)}{E-E_n^{(\alpha)}}
\nonumber\\ &\equiv&\mathcal{R}_{ij}(E)\,\bes{i}(p_ir_0)\,\bes{j}(p'_jr_0)\;,
\label{veff}
\end{eqnarray}
where the RSE propagator $\mathcal{R}_{ij}(E)$ contains an infinite tower
of $s$-channel bare $q\bar{q}$ states, corresponding to the spectrum of
an, in principle, arbitrary confining potential. Here, $E_n^{(\alpha)}$ is
the discrete energy of the $n$-th recurrence in $q\bar{q}$ channel $\alpha$,
$g_i^{(\alpha)}(n)$ \cite{ZPC21p291} is the corresponding coupling to the
$i$-th meson-meson channel, $\bes{i}(p_i)$ is the $L_i$-th order spherical
Bessel function, $p_i$ is the relativistically defined off-shell relative
momentum in meson-meson channel $i$, $r_0$ is a distance parameter, and
$\lambda$ is an overall coupling constant. Note that the spherical Bessel
function results from assuming $^{3\!}P_0$ $q\bar{q}$ pair
creation/annihilation only to take place at a certain distance
$r_0$ \cite{AOP324p1620}.  The closed-form off-energy-shell
$\mathcal{T}$-matrix then reads 
\begin{eqnarray}
\lefteqn{\tmat{i}{j}(p_i,p'_j;E)=} \\[-2mm] && \!\!\!\!
-2\lambda^2r_0\,\sqrt{\mu_ip_i\mu'_jp'_j}\:\bes{i}(p_ir_0)\sum_{m=1}^{N}
\rse_{im}(E)\,\left\{[\One-\Omega\,\mathcal{R}]^{-1}\right\}_{\!mj}\,
\bes{j}(p'_jr_0)\;, \nonumber \\[-5mm] \nonumber
\label{tfinal}
\end{eqnarray}
where
\begin{equation}
\Omega(E) \; = \; -2i\lambda^2r_0\:\mbox{diag}\left(\bes{n}(k_nr_0)\,
\han{n}(k_nr_0)\right) \; ,
\label{loop}
\end{equation}
with $\han{n}(k_nr_0)$ the spherical Hankel function of the first kind,
and $k_n$ the on-shell relative momentum in meson-meson channel $n$.
Finally, the corresponding unitary and symmetric (on-shell)
$\mathcal{S}$-matrix is given by
\begin{equation}
\mathcal{S}_{ij}^{(L_i,L_j)}(k_i,k_j;E)\;=\;\delta_{ij}\:+\:2i
\tmat{i}{j}(k_i,k_j;E)\;.
\end{equation}
Note that the $\mathcal{S}$-matrix is only unitary for real $k_i$, and so
for real meson masses in channel $i$.

\section{Redefining the $\mathcal{S}$-matrix} 
As mentioned above, if we take a complex mass in any of the meson-meson
channels, the $\mathcal{S}$-matrix ceases to be unitary, but stays symmetric.
This requires a redefinition of the $\mathcal{S}$-matrix. Now, note that
\begin{equation}
\mathcal{S}^\dagger \mathcal{S}\equiv A
\label{nonunitary}
\end{equation}
is not unity anymore, but it is still a Hermitian matrix, by definition,
and so with real eigenvalues, which moreover are all positive in this special
case.  Thus, $A$ can be diagonalised by a unitary matrix $U$: 
\begin{equation}
A_d=U\mathcal{S}^\dagger \mathcal{S}U^\dagger \; .
\label{diag}
\end{equation}
Define now
\begin{equation}
\tilde{\mathcal{S}}\equiv \mathcal{S}U^\dagger A_d^{-1/2}U\;,
\label{stilde}
\end{equation}
where
$A_d^{-1/2}$ is real, because of the positive eigenvalues.
Then, it is straightforward to show that
\begin{equation}
\tilde{\mathcal{S}}^\dagger\tilde{\mathcal{S}}=\tilde{\mathcal{S}}
\tilde{\mathcal{S}}^\dagger=\One \; .
\label{sunitary}
\end{equation}
It is not so easy to prove that $\tilde{\mathcal{S}}$ is also symmetric,
but this has been checked numerically with a precision of better than one
part in a trillion.  

\section{Preliminary results}
\begin{figure}[t]
\begin{center}
\epsfysize=250pt
\epsfbox{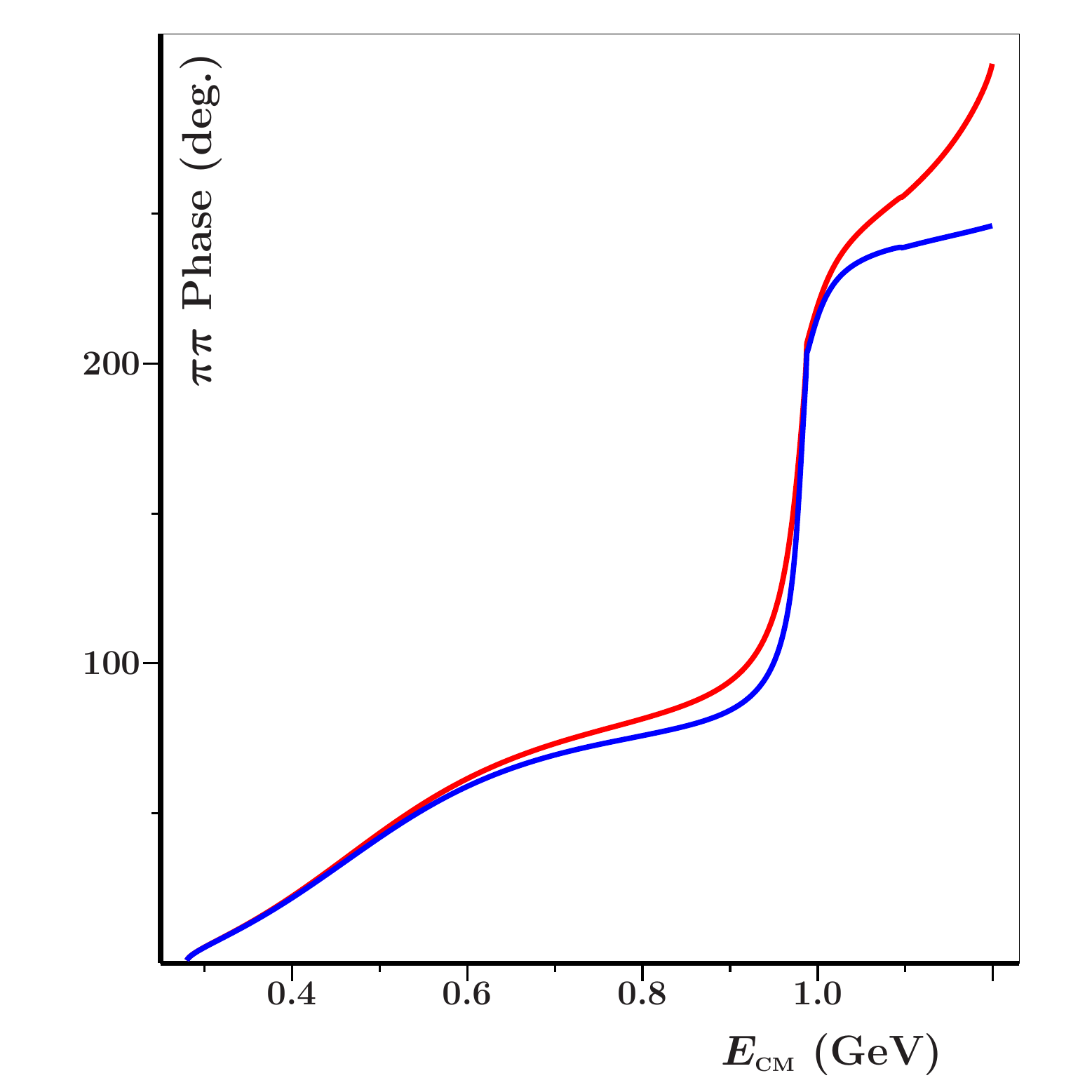}
\end{center}
\mbox{ } \\[-1.5cm]
\caption{RSE calculation of $S$-wave $\pi\pi$ phase shifts. Red (upper) curve:
from standard $S$-matrix and real $\sigma$, $\kappa$, and $\rho$ masses; blue
(lower) curve: from redefined $\mathcal{S}$-matrix in Eq.~(\ref{stilde}) and
complex masses.}
\label{realcomplex}
\end{figure}

Let us now use the redefined $\mathcal{S}$-matrix of Eq.~(\ref{sunitary})
in a purely comparative calculation. Starting point is our \cite{APPPS2p437}
application of the RSE to $S$-wave $\pi\pi$ scattering, with
pseudoscalar-pseudoscalar, vector-vector, and scalar-scalar channels included.
Then, we substitute complex values for the masses of the broadest resonances
in some of these channels, namely for the $f_0(600)$ (alias $\sigma$),
$K_0^\ast(800)$ (alias $\kappa$), and $\rho$ meson. Note that the precise
values we take for the complex masses of the very broad $\sigma$ and $\kappa$
resonances are not so important here, but in future model fits to the data one
should use the best available, ``world-average'' values for the corresponding
pole positions. In Fig.~\ref{realcomplex}, we plot the curves for the two
cases. Quite significant differences become evident, especially above 1~GeV.
Nevertheless, the curve obtained with $\tilde{\mathcal{S}}$ in
Eq.~(\ref{stilde}) and complex masses looks reasonable, though no refit has
been done.

\section{Summary}
In the present study, we have carried out an empirical procedure in 
order to restore unitarity of a coupled-channel $\mathcal{S}$-matrix with
complex masses in the asymptotic states. Although no rigorous mathematical
justification is given here for the purely algebraic transformation of the
original $\mathcal{S}$-matrix, preservation of the mandatory symmetry of
$\mathcal{S}$ gives us confidence that the method makes sense. A direct
comparison of $S$-wave $\pi\pi$ phase shifts, first calculated with the
original $\mathcal{S}$ and using real masses for the $\sigma$, $\rho$, and
$\kappa$ resonances in some of the coupled channels, and then with
the $\tilde{\mathcal{S}}$ of Eq.~(\ref{stilde}) and complex values for the
latter masses, shows significant yet reasonable changes. Nevertheless, our
procedure will have to be tested in concrete fits to the data so as to find
out whether it is really a promising new approach to multiparticle hadronic
decays.  
\section*{Acknowledgements}
We thank the organisers for another exciting and very pleasant workshop.
This work received financial support from the
{\it Funda\c{c}\~{a}o para a Ci\^{e}n\-cia e a Tecnologia}
\/of the {\it Minist\'{e}rio da Ci\^{e}ncia, Tecnologia e Ensino Superior}
\/of Portugal, under contract CERN/FP/83502/2008, and by Instituto Superior
T\'{e}cnico, through fellowship no.\ SFA-2-91/CFIF.

\end{document}